\documentclass[epjc]{sn-jnl}

\usepackage{graphicx}      
\usepackage{amsmath}       
\usepackage{amsfonts}      
\usepackage{amssymb}       
\usepackage{bm}            
\usepackage{siunitx}       
\usepackage{booktabs}      
\usepackage{array}         
\usepackage{caption}       
\usepackage{subcaption}    
\usepackage{hyperref}      
\usepackage{upgreek}       
\usepackage{adjustbox}     
\usepackage{multirow}      
\usepackage{tikz}          
\usetikzlibrary{shapes.geometric, arrows, positioning}

\tikzstyle{startstop} = [rectangle, rounded corners, minimum width=3cm, minimum height=1cm, text centered, draw=black, fill=red!30]
\tikzstyle{process} = [rectangle, minimum width=3cm, minimum height=1cm, text centered, draw=black, fill=orange!30]
\tikzstyle{decision} = [diamond, minimum width=3cm, minimum height=1cm, text centered, draw=black, fill=green!30]
\tikzstyle{preprocess} = [rectangle, rounded corners, minimum width=3cm, minimum height=1cm, text centered, align=center, draw=black, fill=blue!10]
\tikzstyle{arrow} = [thick,->,>=stealth]

\begin{document}

\title[Gamma Neutron Identification in WCDs]{Gamma Neutron Radioactive Source Identification in Water Cherenkov Detectors}

\author[1,2]{A. Núñez Selin}
\author[3,4]{C. Sarmiento Cano}
\author[5]{H. Asorey}
\author[1,4.6,*]{I. Sidelnik}

\affil[1]{Departamento de Física de Neutrones, Centro Atómico Bariloche, Av. Bustillo 9500, S. C. Bariloche, 8400, Argentina}
\affil[2]{Instituto Balseiro, CNEA-UNCuyo, Argentina}
\affil[3]{Escuela de Física, Universidad Industrial de Santander, Bucaramanga, Colombia}
\affil[4]{Departamento de Ciencias Básicas, Universidad Autónoma de Bucaramanga, Bucaramanga, Colombia}
\affil[5]{Fractalis Adventures \& piensas.xyz, Las Rozas Innova, Jacinto Benavente 2, 28232 Las Rozas de Madrid, Spain}
\affil[6]{Consejo Nacional de Investigaciones Científicas y Técnicas (CONICET), Argentina}

\affil[*]{Corresponding author: \href{mailto:ivansidelnik@cnea.gob.ar}{ivansidelnik@cnea.gob.ar}}

\abstract{
Water Cherenkov Detectors (WCDs) are a robust technology widely used in astrophysics, high energy physics, and recently nuclear security applications. They detect high energy interactions through the Cherenkov light emitted by charged particles traveling faster than the speed of light in water.

In this work, we demonstrate the feasibility of gamma-neutron discrimination in WCDs using a combined methodology that integrates statistical analysis with machine learning techniques. The experimental setup employs different shielding configurations to isolate gamma and neutron contributions from a \textsuperscript{241}AmBe source, while \textsuperscript{60}Co and \textsuperscript{137}Cs sources are used to establish a signal to energy calibration.

A statistical analysis based on a $3\sigma$ significance criterion is used to define energy thresholds, enabling a linear relationship between the measured charge spectrum and the deposited energy. Building on this calibration, pulse shape information is further exploited through machine learning methods to improve event classification. An ensemble model based on a soft-voting strategy combining a Bagging classifier, CatBoost, and a Multilayer Perceptron was trained on detector signals acquired under different shielding conditions, achieving an accuracy of 0.816 and an area under the receiver operating characteristic (ROC) curve (AUC) of 0.921.

The combined approach demonstrates that statistical thresholding provides a physically grounded discrimination baseline across the full energy range, while machine learning enhances classification performance at higher energies by leveraging pulse level information. This integrated strategy improves radiation identification capabilities in water Cherenkov detectors, with potential applications in nuclear security and radiation detection.
}


\maketitle

\section{Introduction}
\label{sec1}

Water Cherenkov Detectors (WCDs) are based on the fundamental physics of charged particle interactions and provide robust radiation detection capabilities. When relativistic particles traverse water at velocities exceeding the medium's Cherenkov threshold, they generate distinctive conical photon emissions through the Cherenkov mechanism \cite{frank1991coherent}. This physical principle underpins their widespread adoption in both fundamental physics research as demonstrated by large scale installations like Latin American Giant Observatory (LAGO) \cite{lago_project_2019}, Super Kamiokande \cite{fukuda2003super} and the Pierre Auger Observatory \cite{auger_observatory_2015} and applied nuclear security applications \cite{sidelnik2020enhancing}.

Recent developments have renewed interest in WCDs for radiation monitoring scenarios requiring neutron--gamma discrimination \cite{sidelnik_neutron_detection_2017}. Conventional systems based on \textsuperscript{3}He proportional counters face increasing limitations due to global supply constraints \cite{sachetti_3he_free_2015}, motivating the exploration of alternative detection technologies. In this context, WCDs offer several practical advantages, including scalability, chemical safety, and the ability to exploit temporal and spectral features of detector pulses for particle identification \cite{torres2024enhanced, watanabe_neutron_tagging_2009}.

This work addresses two critical challenges in operationalizing water based detectors for security applications: (1) establishing reliable energy deposition thresholds for coarse particle type discrimination, and (2) developing robust pulse level classification methods for mixed radiation fields. Our experimental framework employs a \textsuperscript{241}AmBe neutron gamma source alongside pure gamma emitters (\textsuperscript{60}Co and \textsuperscript{137}Cs) under controlled shielding configurations. Lead and paraffin shielding geometries enable isolation of neutron and gamma components through differential attenuation properties \cite{chichester2007radiation}.

The methodology combines fundamental detector physics with machine learning refinement. We first establish statistical discrimination thresholds through rigorous calibration of the equivalent charge spectrum, applying significance testing ($3\sigma$ criterion, a conventional choice in radiation detection to ensure strong background rejection while maintaining statistical robustness \cite{eadie1973statistical}) to determine optimal energy cutoffs. Subsequently, detector output pulses for the \textsuperscript{241}AmBe source acquired under various shielding configurations are analyzed using heterogeneous machine learning models. The ensemble approach strategically combines models with complementary prediction correlation to maximize classification robustness.

This study builds upon our previous work on neutron detection with water Cherenkov detectors \cite{sidelnik_neutron_detection_2017,sidelnik_neutron_capabilities_2018,sidelnik2020enhancing}, as well as more recent developments reported in \cite{selin2026two,betancourt_rich_2025}.

The main contributions of this work are summarized as follows:

\begin{itemize}
\item A systematic calibration procedure linking deposited energy to incident radiation characteristics in pure water WCDs  
\item A two stage discrimination framework combining statistical energy thresholds with machine learning (ML) pulse classification  
\item A quantitative assessment of ensemble model diversity in radiation signal processing. 
\end{itemize}

These contributions directly address operational requirements in nuclear security applications, where discrimination capability must be balanced with robustness and interpretability. The proposed approach preserves the intrinsic advantages of water based detectors while mitigating limitations associated with low energy signal discrimination.

The paper is organized as follows. Section \ref{sec2} describes the experimental setup and data acquisition system. Section \ref{sec3} presents the energy calibration procedure and the statistical discrimination method. Section \ref{sec:2stage} introduces the two-stage gamma--neutron classification approach. Section \ref{sec:ml} details the machine learning pipeline and ensemble optimization. Section \ref{sec5} discusses operational implications and future perspectives for radiation detection systems.

\section{Water Cherenkov Detectors (WCD)}
\label{sec2}

Water Cherenkov Detectors (WCDs) are based on the detection of Cherenkov radiation produced when a charged particle traverses a dielectric medium with
refractive index \(n > 1\) at a speed exceeding the local speed of light, i.e., \(v > c_m = c/n\) (with \(c\) representing the speed of light 
in vacuum). Originally observed by P.E. Cherenkov in 1937 \cite{cherenkov1934visible} and later explained by Frank and Tamm \cite{frank1991coherent}, this effect arises from the coherent emission of electromagnetic radiation triggered by the asymmetric polarization of electrons in the medium. The kinetic energy threshold for Cherenkov radiation emission by electrons is \(\text{T}_{\text{th, water}} = \SI{262}{\kilo\electronvolt}\) \cite{sidelnik2020enhancing}.

\subsection{Detector Description}
\label{subsec:detector}

The experimental setup comprises a Water Cherenkov Detector (WCD) installed at the Laboratorio de Física de Neutrones of the Centro Atómico de Bariloche, Argentina \cite{sidelnik2020enhancing,sidelnik_neutron_capabilities_2018}. The detector is designed with a cylindrical geometry and an active volume of 1.0 m\(^3\) (94 cm in diameter \(\times\) 147 cm in height), optimized for detailed studies of radiation interactions. Some of the key components of the system include:

\begin{itemize}
    \item \textbf{Active Medium:} Pure water, which functions simultaneously as the Cherenkov radiator and neutron moderator.
    \item \textbf{Photodetection System:} A Photonics XP1802 9'' photomultiplier tube (PMT) \cite{wright2017photomultiplier}, ensuring single-photon sensitivity across the Cherenkov emission band (300--650 nm).
    \item \textbf{Signal Acquisition:} A Red Pitaya STEMLAB 125-14 FPGA board \cite{Pitaya125} sampling at 125 MS/s, integrated with the LAGO DAQ ecosystem \cite{LAGO_rep} for precise waveform digitization.
    \item \textbf{Voltage Supply:} A LAGO voltage control board that provides regulated bias voltages to the PMT base.
    \item \textbf{Optical Reflector:} The interior surfaces are lined with 0.12 mm Tyvek\textregistered \cite{filevich_tyvek_reflectivity_1999} to enhance light collection through diffuse reflection.
\end{itemize}

The PMT is mounted on the detector lid, positioned at the center of the top surface and aligned along the cylindrical axis. Its high voltage is regulated via the Latin American Giant Observatory control board, which concurrently monitors the system through the Red Pitaya interface, while the Tyvek lining maximizes photon collection efficiency.

\subsection{Detector Response}
\label{subsec:response}

In WCDs, radiation interactions are recorded as current pulses, with the charge integrated under each pulse encapsulating vital information about the incident radiation \cite{sidelnik2020enhancing}. The primary interactions in the detection medium are:

\begin{itemize}
    \item \textbf{Gamma Radiation:} Interacts through the Compton effect, photoelectric effect, or pair production (for energies exceeding 1.022 MeV). Owing to the large active volume of the detector, the probability of gamma interaction is significantly enhanced, increasing the likelihood of full or partial energy deposition within the medium. Secondary electrons generated by these processes produce Cherenkov photons that are subsequently detected by the PMT.
    \item \textbf{Charged Particles:} Directly generate Cherenkov photons via ionization processes, provided their energy surpasses the emission threshold, or contribute via bremsstrahlung.
    \item \textbf{Neutron Radiation:} Neutrons are moderated by water and primarily captured by hydrogen nuclei (\(^{1}\)H), leading to the emission of 2.22 MeV gamma rays which then interact with the medium.
\end{itemize}

\subsection{Radioactive Sources}
\label{subsec:sources}

Three distinct radioactive sources were employed in the experiments. Each source was positioned at the detector's average height and placed 30 cm away from the active volume, measurements were conducted for 5 minutes for each source. The emission rates quoted below correspond to the source activities at the time of measurement:

\begin{itemize}
    \item \textbf{\(^{60}\)Co:} Emits gamma rays at 1.17 MeV and 1.33 MeV \cite{knoll2010radiation}, with an emission rate of \(1.42 \times 10^5\) gammas/s.
    \item \textbf{\(^{137}\)Cs:} Emits gamma rays at 0.66 MeV \cite{knoll2010radiation}, with an emission rate of \(2.78 \times 10^5\) gammas/s.
    \item \textbf{\textsuperscript{241}AmBe:} Produces 4.44 MeV gamma rays \cite{beckurts2013neutron} and a continuous spectrum of neutrons \cite{geiger1964neutron}, with a total emission rate of \(2.04 \times 10^6\) neutrons/s.
\end{itemize}

\subsection{Shielding Configurations}
\label{subsec:shield}

To effectively distinguish the various radiation components from the \textsuperscript{241}AmBe source, several shielding configurations were implemented:

\begin{itemize}
    \item \textbf{Lead (10 cm):} Attenuates 4.44 MeV gamma rays by approximately two orders of magnitude (\(\sim 99\%\)), significantly reducing their contribution while still allowing a small residual component due to incomplete attenuation and secondary photon buildup.

    \item \textbf{Borated Paraffin (10 cm) + Cadmium Sheet (0.1 cm):} The paraffin efficiently moderates fast neutrons to thermal energies through elastic scattering with hydrogen, while the boron content provides strong absorption of thermalized neutrons via the \(^{10}\)B(n,\(\alpha\))\(^{7}\)Li reaction. This configuration suppresses more than 90\% of the neutron flux. The cadmium layer acts as a final absorber for residual thermal neutrons due to its high capture cross section, ensuring that the detected signal is dominated by the 4.44 MeV gamma component.

    \item \textbf{Lead (10 cm) + Pure Paraffin (15 cm):} Combines gamma attenuation with neutron moderation. The lead reduces the 4.44 MeV gamma flux by roughly two orders of magnitude, while the paraffin thermalizes neutrons, which are subsequently captured predominantly by hydrogen nuclei. This capture process produces 2.22 MeV gamma rays, primarily generated outside the detector volume.
\end{itemize}

\section{Energy Level Classification}
\label{sec3}

In conventional detectors capable of operating as spectrometers, the calibration of energy channels is typically carried out using standardized radioactive sources of different origins. However, in the case of WCDs, direct access to the deposited energy of individual events is not straightforward, and the estimation of event energies generally relies on indirect analysis techniques. Among the most commonly employed methods are the identification of the muon hump and the use of the Michel electron spectrum \cite{otiniano2022,sarmiento2022arti}.

In the present work, we employ distinct radiation sources that produce characteristic and distinguishable signatures in the detector charge spectrum. Based on these signatures, we introduce a statistical quantity that explicitly incorporates the charge channel information when the signal can be clearly discriminated from the background. This approach allows for a consistent comparison between different radiation sources and detector responses. To implement this methodology, we define a \textit{cutoff point}, which serves as a reference threshold for separating signal dominated regions from background dominated ones in the charge spectrum.

\subsection{Cutoff Point Determination}

The \textit{cutoff point} (CP) is defined as the first histogram bin at which the source spectrum becomes statistically indistinguishable from the background. In the context of the gamma spectra analysis, this bin corresponds to the maximum energy deposition fully contained within the detector, enabling the calibration between charge in analogic to digital units (ADU) and incident gamma energy (MeV).

The cutoff is determined using a $3\sigma$ compatibility criterion. Since source and background measurements were acquired with identical acquisition times and follow independent Poisson statistics, the standard deviation of the bin by bin difference is

\begin{equation}
\sigma = \sqrt{N_{\mathrm{src}} + N_{\mathrm{bkg}}}.
\label{eq:sigma_total}
\end{equation}

The cutoff bin is defined as the first bin satisfying

\begin{equation}
\left|N_{\mathrm{src}} - N_{\mathrm{bkg}}\right| < 3\sigma.
\label{eq:3sigma}
\end{equation}

A detailed treatment of the associated uncertainty in the cutoff position is presented in Appendix~\ref{app:cutoff_uncertainty}.

\begin{figure}
    \centering
    \begin{subfigure}{0.48\textwidth}
        \centering
        \includegraphics[width=\linewidth]{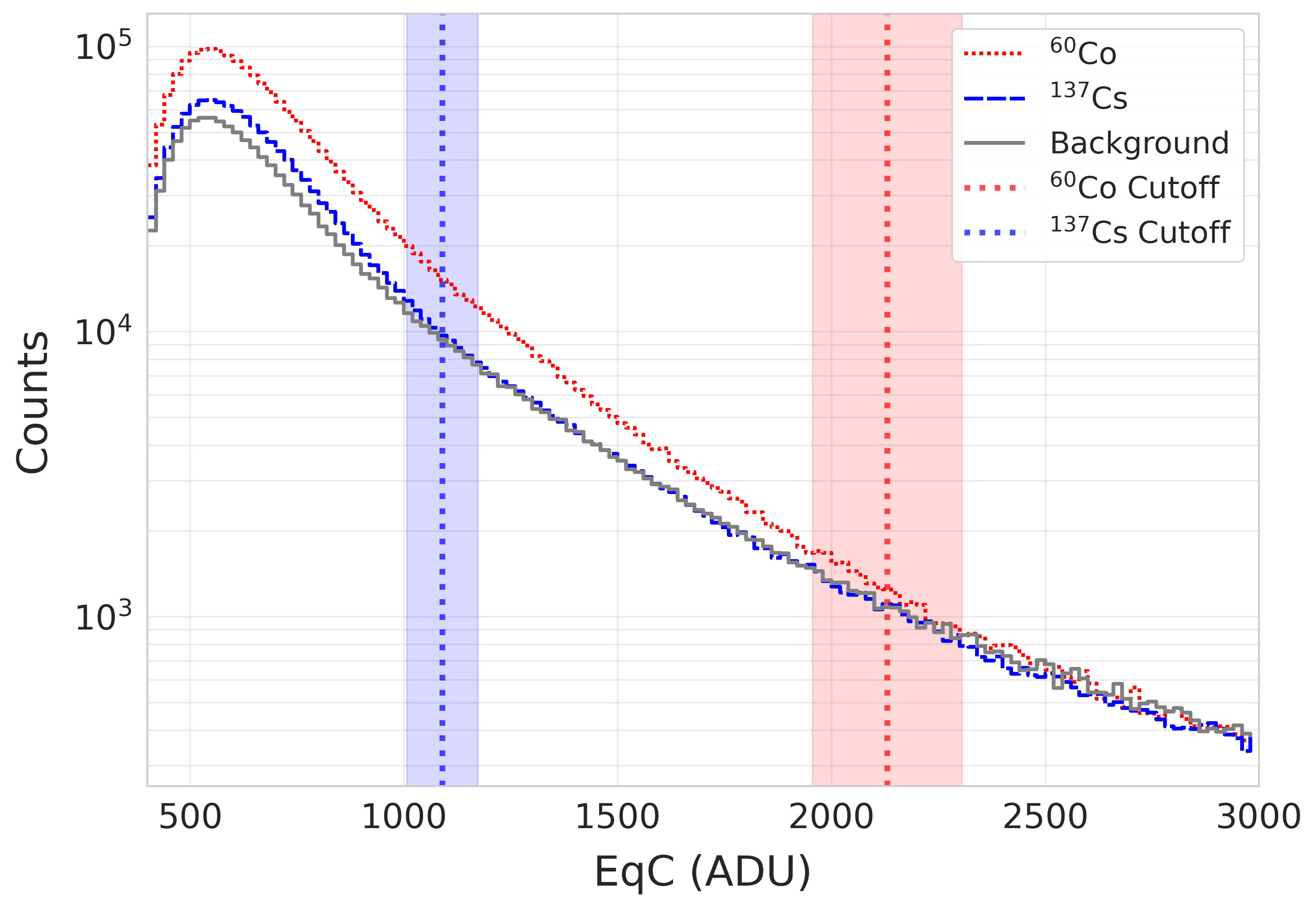}
        \caption{}
    \end{subfigure}
    \begin{subfigure}{0.48\textwidth}
        \centering
        \includegraphics[width=\linewidth]{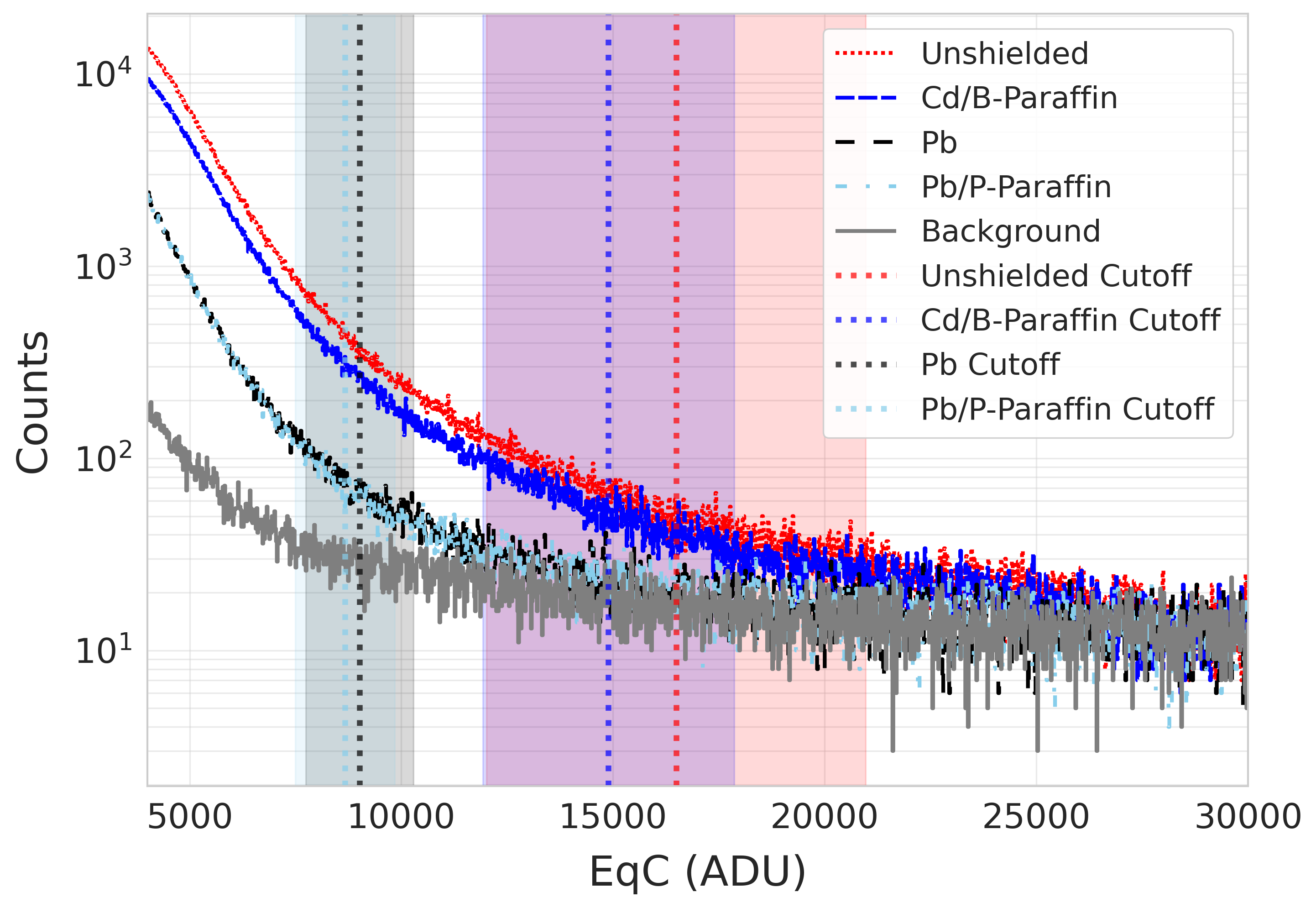}
        \caption{}
    \end{subfigure}
    \caption{Equivalent charge spectra (ADU) with \(3\sigma\) cutoff markers. Colored bands show total uncertainty (\(\sigma_\mathrm{cutoff}\)), obtained from the quadratic combination of statistical, threshold-sensitivity, and discretization contributions. Left hand side: Charge spectra for \textsuperscript{137}Cs (0.66 MeV) and \textsuperscript{60}Co (1.17 + 1.33 MeV). Right hand side: Charge spectra for \textsuperscript{241}AmBe under shielding configurations.}
    \label{fig:cutoff_spectra}
\end{figure}

Figure~\ref{fig:cutoff_spectra} presents the charge spectra for gamma sources (\textsuperscript{137}Cs and \textsuperscript{60}Co, in the left) as well as for mixed neutron gamma emissions from \textsuperscript{241}AmBe (right) under different shielding configurations (as can be seen in Sec.~\ref{subsec:shield}). The vertical dashed lines denote the \(3\sigma\) cutoff points, as defined in Eq.~\ref{eq:3sigma}, while the colored bands illustrate the total uncertainty. A quantitative summary of these cutoff values is provided in Table~\ref{tab:calibration}.

The identification of the cutoff point relies on achieving a \(3\sigma\) statistical significance between source and background counts. Consequently, the ability to resolve the cutoff depends on the total number of detected events, which is governed by the source activity, the source-detector distance, and the acquisition time. For weak sources or larger distances, longer acquisition times may be required to reach the same level of statistical confidence. In the present laboratory conditions, where the sources were positioned close to the detector and exhibit relatively high activity, all measurements were performed over 5 minutes, ensuring sufficient statistics for a robust determination of the cutoff. Shorter acquisition times (e.g., 1 minute), as in dynamic scenarios such as moving sources, would reduce the statistical significance and could hinder reliable cutoff identification unless compensated by higher count rates.

\begin{table}
\centering
\adjustbox{max width=\linewidth}{%
\begin{tabular}{>{\centering\arraybackslash}m{3cm} >{\centering\arraybackslash}m{2.5cm} >{\centering\arraybackslash}m{2cm} >{\centering\arraybackslash}m{3cm}}
\toprule
\textbf{Source} & \textbf{Radiation Type} & \textbf{Energy (MeV)} & \textbf{Cutoff Point (ADU)} \\
\midrule
\({}^{137}\text{Cs}\) & \(\gamma\) & 0.66 & \(1090 \pm 80\) \\
\({}^{60}\text{Co}\)  & \(\gamma\) & 1.17, 1.33 & \(2130 \pm 180\) \\
\({}^{241}\mathrm{AmBe}\) (Pb) & n & 2.22 & \(9000 \pm 1300\) \\
\({}^{241}\mathrm{AmBe}\) (Pb/P-Paraffin) & \(\gamma\) & 2.22 & \(8700 \pm 1200\) \\
\({}^{241}\mathrm{AmBe}\) (Cd/B-Paraffin) & \(\gamma\) & 4.44 & \(14900 \pm 3000\) \\
\({}^{241}\mathrm{AmBe}\) (Unshielded) & \(\gamma\) + n & 2.22, 4.44 & \(16500 \pm 4500\) \\
\bottomrule
\end{tabular}%
}
\caption{Cutoff energy values for different radioactive sources. The close agreement between the Pb and Pb/P-Paraffin configurations (\(8700 \pm 1200\) vs. \(9000 \pm 1300\)) supports the hypothesis that neutrons, once thermalized and absorbed in the detector, predominantly produce a secondary 2.22 MeV gamma. Likewise, the similarity between the Cd/B-Paraffin (\(14900 \pm 3000\)) and Unshielded (\(16500 \pm 4500\)) configurations suggests that the cutoff is primarily determined by the highest energy event (4.44 MeV gamma emission).}
\label{tab:calibration}
\end{table}

With this method, a source with a cutoff energy below the neutron threshold of \(9000 \pm 1300\) ADU can be identified as a pure gamma emitter for this detector configuration. This is because, regardless of the presence of neutron emitting sources, the detector's active medium thermalizes and absorbs neutrons, thereby generating a secondary 2.22 MeV gamma. The data in Table~\ref{tab:calibration} confirm this behavior: the cutoff values for the Pb/P-Paraffin configuration (representing an incident 2.22 MeV gamma generated externally) are similar to those for the Pb configuration (\(8700 \pm 1200\) vs. \(9000 \pm 1300\)). Furthermore, the cutoff values for the Cd/B-Paraffin (\(14900 \pm 3000\)) and Unshielded (\(16500 \pm 4500\)) configurations indicate that the dominant contribution arises from the highest energy event, namely the 4.44 MeV gamma emission. Thus, the calibration cutoff is primarily determined by the most energetic gamma interaction rather than by lower energy components. As a reference scale, the muon hump associated with nearly vertical through-going muons is located at approximately \(7 \times 10^{4}\) ADU, well above the cutoff values reported here.

\subsection{Energy Calibration}

\begin{figure}
    \centering
    \begin{subfigure}{0.48\textwidth}
        \centering
        \includegraphics[width=\linewidth]{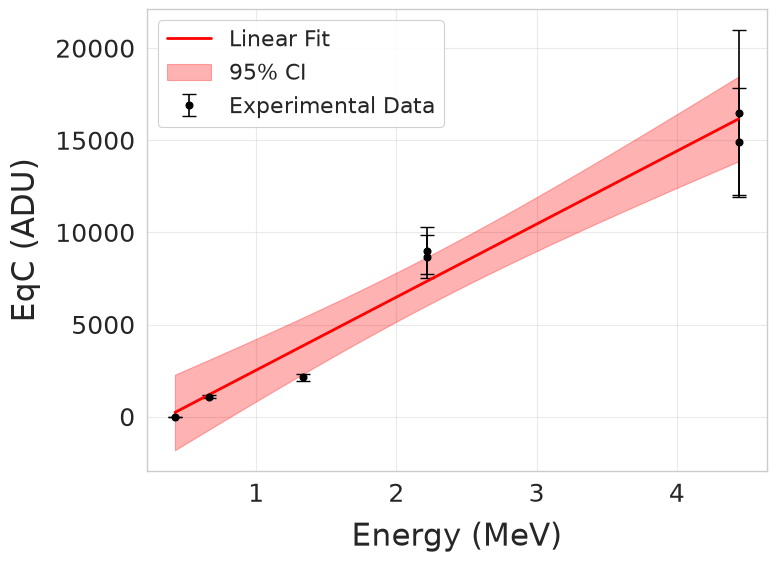}
        \caption{}
    \end{subfigure}
    \begin{subfigure}{0.48\textwidth}
        \centering
        \includegraphics[width=\linewidth]{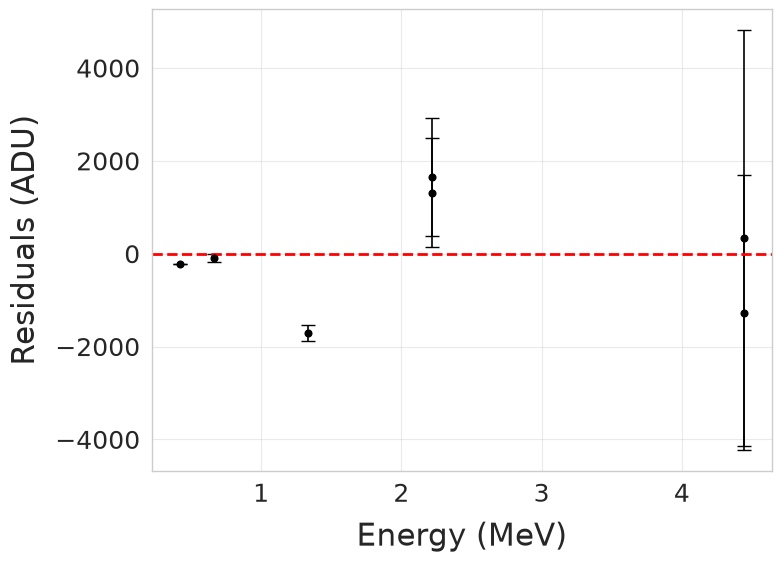}
        \caption{}
    \end{subfigure}
    \caption{Left: Energy calibration curve showing the measured data with error bars (black), the best fit linear regression line (red), and the 95\% CI band (red band). Right: Residuals relative to the linear fit, with error bars representing cutoff determination uncertainties.}
    \label{fig:calibration}
\end{figure}

The measurements presented in Fig.~\ref{fig:cutoff_spectra} are used to establish a calibration between the recorded charge (ADU) and the maximum characteristic gamma ray energy emitted by each source (MeV). Figure~\ref{fig:calibration} shows the calibration results obtained from the experimental data reported in Table~\ref{tab:calibration}. The Cherenkov threshold for electrons in water ($\text{T}_e = 0.262$ MeV) corresponds to an incident gamma ray energy of approximately $\text{E}_\gamma = 0.421$ MeV, for which the Compton edge reaches this threshold. This value was therefore used as the zero charge reference of the empirical incident energy scale.

For practical applications, the inverse relation allowing the estimation of deposited energy from measured charge is given by:

\begin{equation}
    \text{E} \, (\text{MeV}) = (3.20 \pm 0.16)\times 10^{-4} \, \text{EqC (ADU)} + 0.42,
    \label{eq:fit_inverse}
\end{equation}

which is valid for this detector and configuration.

The observed linear trend indicates that the detector response is approximately proportional to the deposited energy within the explored range. The residuals do not exhibit a clear systematic bias; however, their associated uncertainties increase significantly at higher energies. This behavior is expected, as the number of detected events decreases with increasing energy, leading to poorer counting statistics.

In this regime, the uncertainty associated with the threshold selection procedure (see Appendix~A) becomes dominant, contributing approximately 80\% of the total uncertainty. As a result, the spread in the residuals is primarily driven by the growth of the uncertainty bars rather than by large deviations of the data points from the model.

Consequently, the reduced \(\chi^2\) exceeds unity, indicating that the uncertainties are not purely statistical but are significantly influenced by systematic effects inherent to the cutoff definition. Therefore, the calibration should be interpreted as an effective mapping for this specific detector setup, where the reliability of the method is better represented by the global uncertainty band of the fit rather than by the precision of individual calibration points.


\section{The ``Two stage gamma neutron'' classification method}
\label{sec:2stage}

\begin{figure*}
    \centering
    \includegraphics[width=0.8\textwidth]{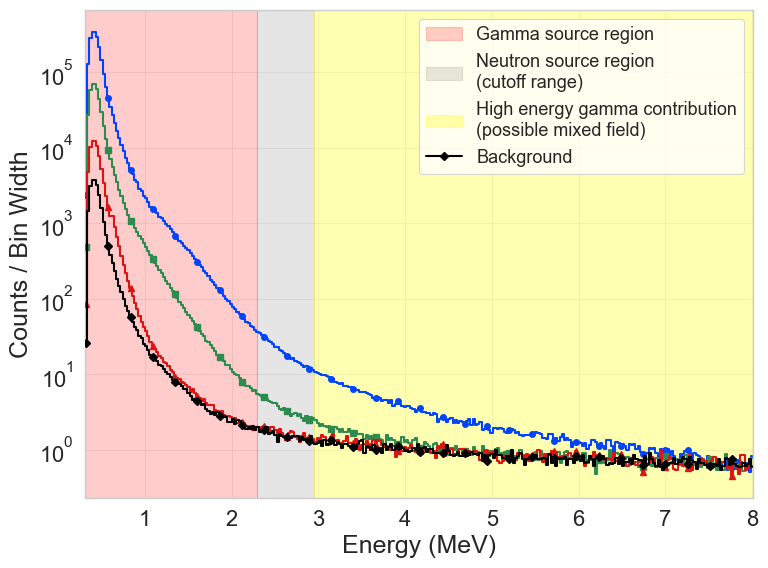}
    \caption{The background-equivalent charge (ADUs) is converted into radiation energy (MeV) using Eq.~\ref{eq:fit_inverse}. The green band represents the neutron detection threshold, marking the transition from red (only gamma, no neutrons) to yellow (possible neutron presence). Three different spectra showing the effect of a pure gamma source (red), a neutron source (green) and a mixed field source (blue) can be seen, altogether with a background measurement (black).}
    \label{fig:twostage}
\end{figure*}

By leveraging the calibration performed for the different sources as shown in section \ref{sec3}, a linear correlation can be established between the incident radiation energy (in MeV) and the equivalent charge bin (in ADUs). This correlation enables the transformation of the equivalent charge spectrum into an incident radiation energy spectrum, as depicted in Fig.~\ref{fig:twostage}. In this figure, an example of the background, in black line, can be seen, as long as three different spectra showing the effect of a pure gamma source (red), a neutron source (green) and a mixed field source (blue). The idea is that in the presence of an unknown source once its cutoff point in this transformed spectrum is identified, it can provide an estimate of the maximum energy of the detected source, resulting in a possible source identification.

\begin{figure*}
    \centering
    \includegraphics[width=0.8\textwidth]{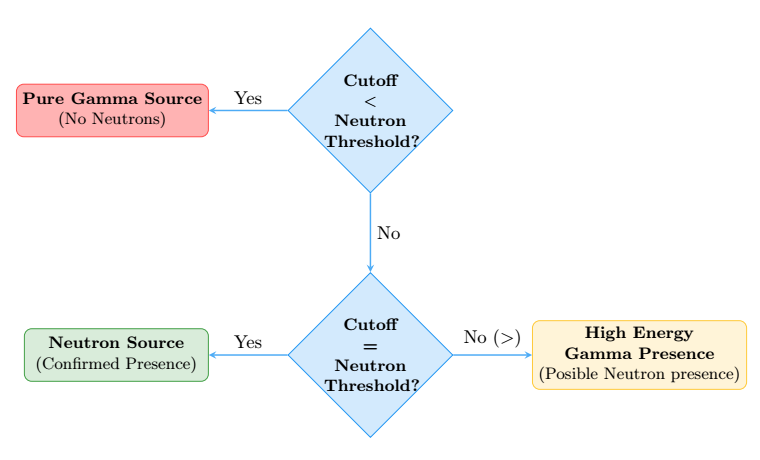}
    \caption{Classification workflow: Decision nodes (blue) route sources to pure gamma (red), confirmed neutron (green), or ambiguous cases (yellow).}
    \label{fig:workflow}
\end{figure*}

The converted energy spectrum can be interpreted as a \textbf{``Traffic light''}, that depending of where the source contribution arises, a classification can be done. In Fig.~\ref{fig:workflow} a graphic representation of the source classification workflow is shown. Briefly, in the presence of an unknown source the \textbf{``Traffic light''} will work like this:

\begin{itemize}
     \item \textbf{Red (Below Threshold):} If the spectral cutoff falls \textit{below} the threshold, the neutrons will be entirely discarded. The source contains only gamma radiation.
    
    \item \textbf{Green (At Threshold):} If the cutoff falls \textit{at} the threshold a neutron presence can be confirmed. Neutrons dominate the spectral endpoint, though coexisting gamma rays (with energies \(\leq\) threshold) may contribute.
    
    \item \textbf{Yellow (Above Threshold):} If the cutoff falls \textit{above} the threshold, there are exclusively high energy gamma rays. Neutrons, if present, exist only at lower energies, below the measured cutoff (at neutron cutoff).
\end{itemize}

In this scheme, it must be noted that the threshold is a \textit{necessary but insufficient condition} for neutron detection: sources above the threshold may contain neutrons (e.g., Unshielded configuration) or not (e.g., Cd/B-Paraffin configuration). To further enhance source discrimination beyond the neutron threshold, a machine learning based classification approach is implemented. A soft-voting ensemble classifier was trained to differentiate neutron and gamma signals at the pulse level. This method will be developed in Section \ref{sec:ml}.

\begin{figure}
    \centering
    \begin{subfigure}{0.48\textwidth}
        \centering
        \includegraphics[width=\linewidth]{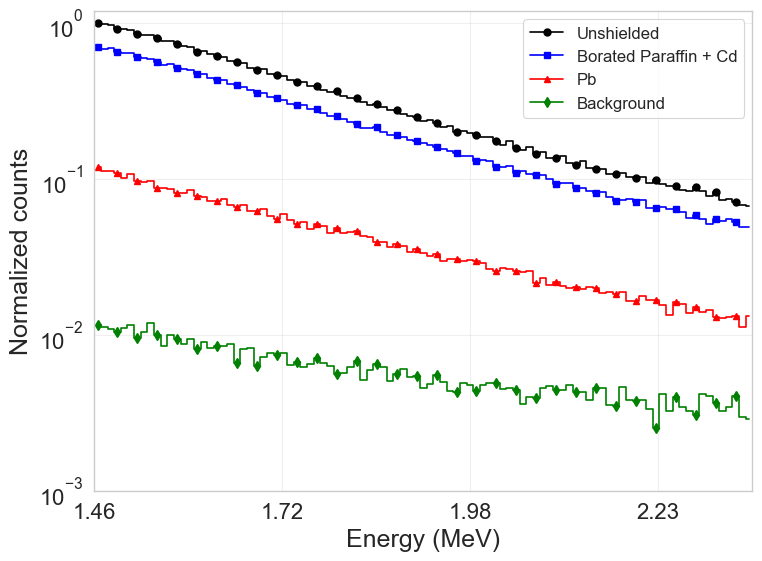}
        \caption{}
    \end{subfigure}
    \begin{subfigure}{0.48\textwidth}
        \centering
        \includegraphics[width=\linewidth]{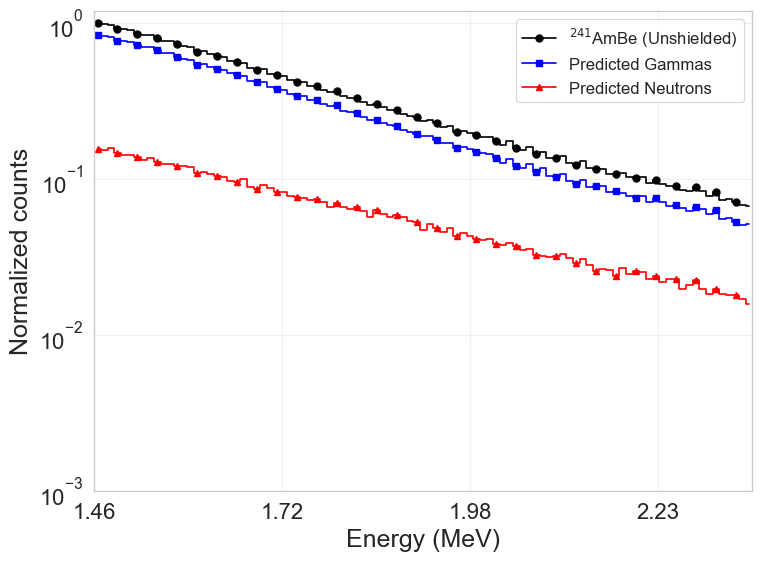}
        \caption{}
    \end{subfigure}
    \caption{Normalized charge spectra (left) and machine learning classification results (right) for neutron and gamma contributions in the detector. The left panel includes experimental measurements under different shielding configurations for the \textsuperscript{241}AmBe source, along with background. The right panel shows the predicted gamma and neutron components obtained by applying the trained model to pulses from the unshielded source, which were not used during training. The experimentally measured unshielded spectrum is also shown for reference. The y-axis is presented on a logarithmic scale to emphasize differences in the spectral shapes.}
    \label{fig:ambe_predicted}
\end{figure}

Figure~\ref{fig:ambe_predicted} illustrates the application of the trained machine learning classifier to decompose the mixed radiation field of an unshielded \textsuperscript{241}AmBe source. The left panel shows the experimentally measured charge spectra under different configurations: the unshielded source (mixed neutron and gamma field), lead shielding (neutron dominated configuration), borated paraffin with cadmium (gamma dominated configuration), and background. All spectra are normalized to the global maximum count in order to enable a direct comparison of their shapes.

The classifier was trained exclusively on labeled data obtained from the borated paraffin + cadmium (gamma dominated) and lead (neutron dominated) configurations. Pulses from the unshielded \textsuperscript{241}AmBe source were not used during training and therefore constitute an independent test of the model.

After training, the classifier was applied to pulses from the unshielded \textsuperscript{241}AmBe source. The right panel presents the predicted neutron and gamma components obtained from this inference step, effectively performing a data driven separation of the mixed radiation field. For comparison, the experimentally measured unshielded \textsuperscript{241}AmBe spectrum is also included in the right panel as a reference, allowing a direct visual assessment of how the predicted components relate to the original mixed signal.

\section{Pulse Level Classification}
\label{sec:ml}

As shown in previous sections, for radiation sources with energies above the neutron emission threshold, it is not immediately evident whether they consist of mixed neutron-gamma emissions or solely high energy gamma radiation. Consequently, a detailed analysis is required to ascertain the true composition of the radiation detected.

The previously established threshold criterion facilitated an initial discrimination of sources below each energy range, where the presence of natural background introduced significant noise into the pulse analysis. This filtering step was crucial for enhancing the accuracy of subsequent evaluations of high energy sources.

It is important to emphasize that the machine learning stage does not attempt to perform high resolution gamma spectroscopy on individual events or replace threshold based screening. Instead, it classifies radiation fields by exploiting statistical differences in pulse shape distributions resulting from multiple Compton interaction topologies within the detector's active volume.

\subsection{Machine Learning and Signal Analysis}

Machine Learning (ML) has become an essential tool in particle physics and radiation detection for processing the large volumes of data generated in experiments, enabling event classification, particle identification, and discrimination between signals of interest and background noise. In Cherenkov detector experiments, for instance, ML techniques are employed to differentiate between particle types, thereby optimizing analysis efficiency and improving the accuracy of event identification \cite{jamieson2022using, tiras2024comprehensive, torres2024enhanced}. Following this approach, our method focuses on distinguishing between the 4.44 MeV gamma emissions from the \(^{241}\)AmBe source, and neutron capture induced signals, all while maintaining a fixed deposited energy range within the detector.

\subsection{Feature Analysis}

Using the measurements shown in Section \ref{sec3} and with the aim of enhance the classification accuracy, we defined a specific charge range, limiting the spectrum to 4500--8000 ADUs. This selection maximizes the neutron signal relative to background noise and minimizes interference from natural radiation. The chosen maximum of 8000 ADUs is near the neutron cutoff value (\(9010 \pm 1271\) ADUs), ensuring that the source signal dominates while background contributions remain at least an order of magnitude lower.

\begin{figure}
    \centering
    \begin{subfigure}{0.48\textwidth}
        \centering
        \includegraphics[width=\linewidth]{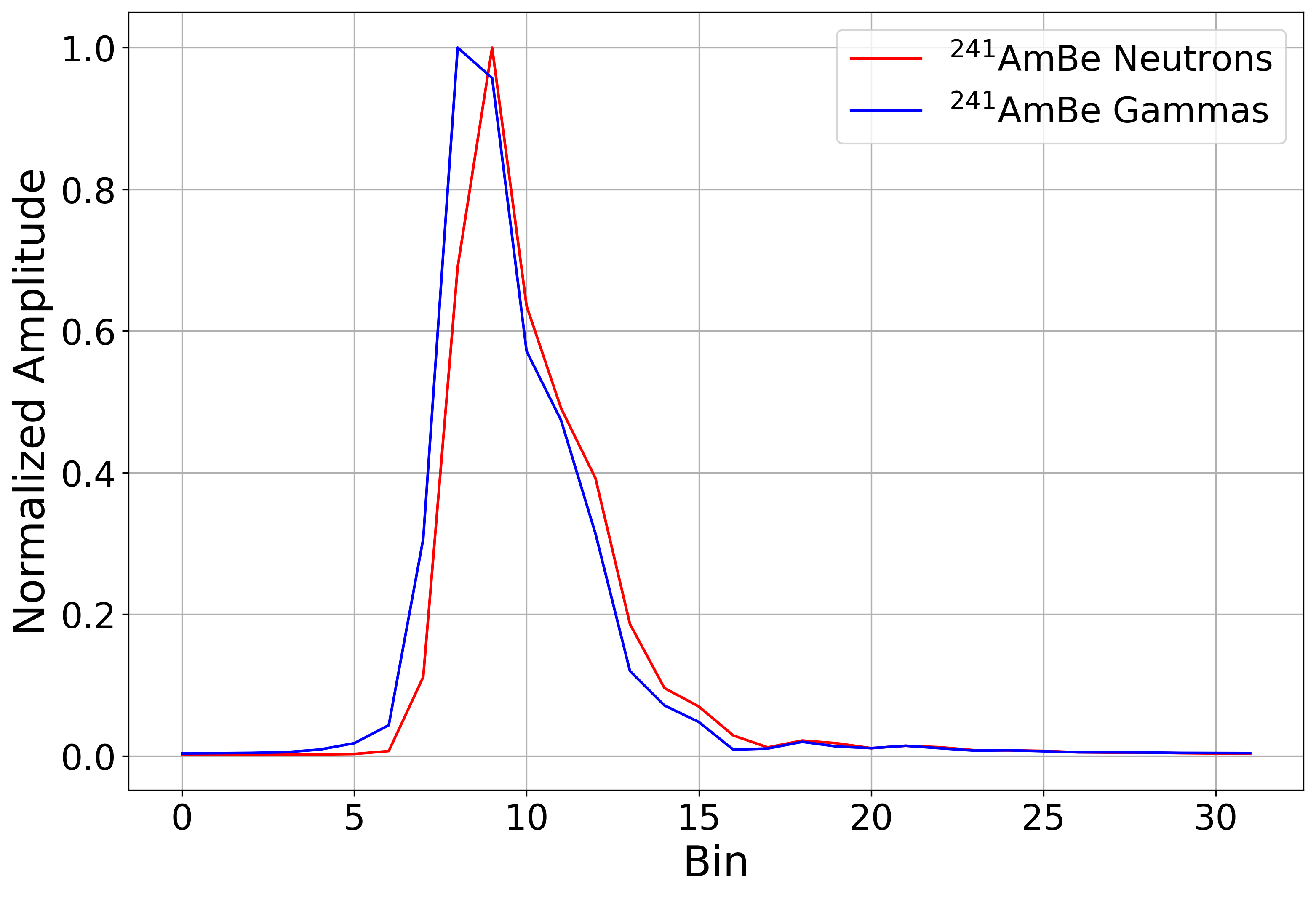}
        \caption{}
    \end{subfigure}
    \begin{subfigure}{0.48\textwidth}
        \centering
        \includegraphics[width=\linewidth]{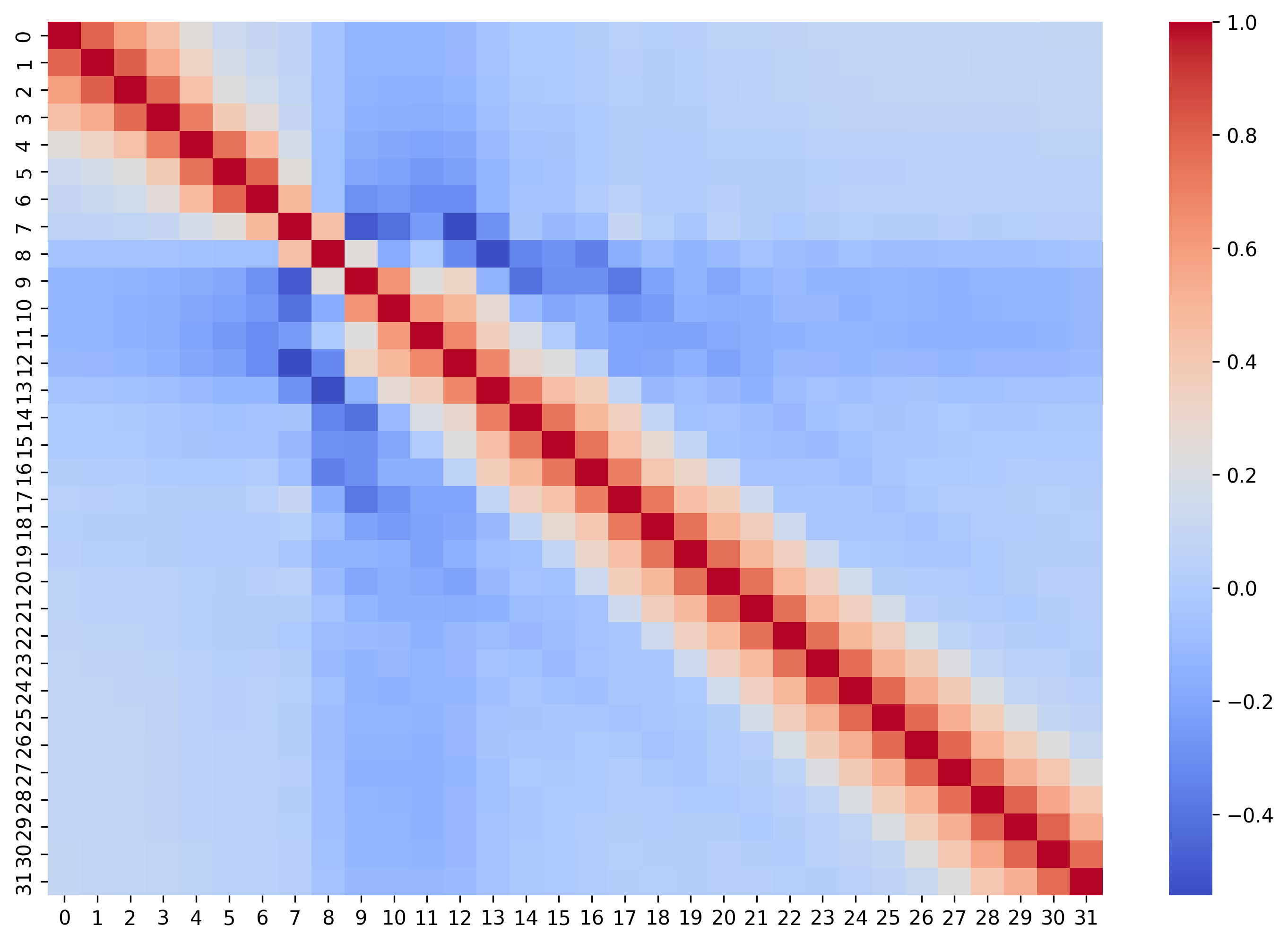}
        \caption{}
    \end{subfigure}
    \caption{Gamma and neutron pulse analysis using the \(^{241}\)AmBe source. Left: Normalized average current pulse. Gamma signals were obtained with Cd/B-Paraffin, while neutron signals were acquired with Pb shielding. Right: Correlation matrix that quantifies the linear relationships between pulse features. The equivalent charge range considered is 4500--8000 ADUs.}
    \label{fig:pulse}
\end{figure}

Each detected pulse comprises 32 temporal bins, where the amplitude of each bin represents the number of charge carriers collected by the photomultiplier tube (PMT) within an 8 ns window, digitized by the ADC of the Red Pitaya. Figure \ref{fig:pulse}, left, shows the normalized average pulse shape for signals acquired with Pb and Cd/B-paraffin shielding.

Most pulse structure development occurs between bins 4 and 17, while the remaining bins exhibit minimal variation. The correlation matrix (Fig. \ref{fig:pulse}, right) shows that the 32 time bins are moderately correlated (\(\lvert\rho\rvert < 0.9\)), indicating sufficient feature independence for multivariate classification. Notably, the anticorrelation between the rise and decay phases suggests these regions contain complementary information for distinguishing signal types.

\subsection{Data Preprocessing Pipeline}

\begin{figure*}
    \centering
    \includegraphics[width=0.7\textwidth]{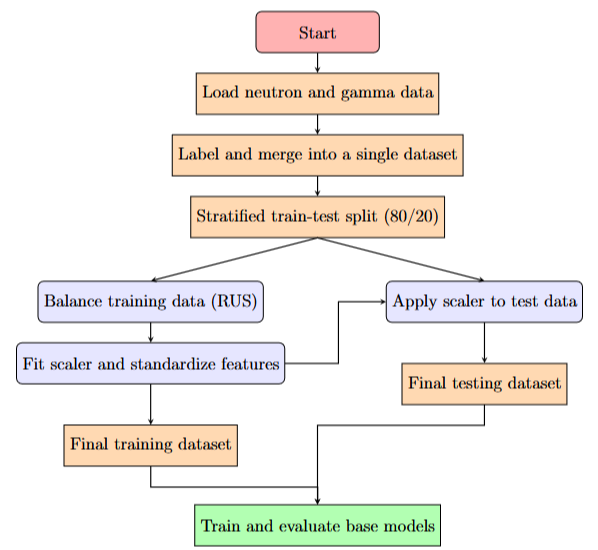}
    \caption{Data preprocessing pipeline for neutron and gamma classification. The left branch processes the training set by applying class balancing and feature scaling, while the right branch ensures consistency in test set transformations. Both branches merge in the final stage, where classification models are trained and evaluated.}
    \label{fig:prepro_pipeline}
\end{figure*}

Preprocessing is a fundamental step to ensure that the classification models yield reliable and unbiased results. The complete workflow of the preprocessing pipeline is illustrated in Figure~\ref{fig:prepro_pipeline}. Each stage is color coded to indicate its role within the pipeline: red denotes the start and end points, orange highlights initial tasks, blue represents auxiliary processing steps, and green marks stages related to the training and evaluation of classification models.

The pipeline begins by loading the raw neutron and gamma data from CSV files. Each sample is labeled (1 for neutrons and 0 for gammas) and combined into a unified dataset. To maintain the original class distribution, the dataset is then split into training (80\%) and testing (20\%) subsets using stratified sampling~\cite{kohavi1995study}. 

To prevent model bias towards the majority class (gamma events), we applied Random Under Sampling (RUS)~\cite{lemaavztre2017imbalanced} to the training set, achieving a balanced 1:1 neutron-gamma ratio (82,967 events each). All features (the 32 time bins) were then standardized using StandardScaler~\cite{pedregosa2011scikit} to ensure they contribute equally to the model training. Critically, the test set (20\% of the data) was kept isolated and was transformed using the parameters learned from the training set to maintain physical validity and unbiased evaluation. To sum up, this pipeline produces training data with balanced classes (1:1 neutron-gamma ratio) while maintaining physically realistic test conditions. 

Using the preprocessed dataset, we trained over 25 classifiers spanning different algorithm families to evaluate their capability to distinguish neutron and gamma signals. The tested models included tree-based ensembles \cite{breiman2001random, chen2016xgboost}, linear models \cite{hastie2009elements}, support vector machines \cite{cortes1995support}, neural networks \cite{goodfellow2016deep}, and distance-based models \cite{bishop2006pattern}. 

All classifiers were evaluated using multiple metrics: accuracy, precision, recall, F1 score, and log loss (where applicable). To focus the comparative analysis on models with meaningful discriminative capability while avoiding redundancy among poorly performing variants of similar algorithms, an accuracy threshold of 0.75 was adopted as a pragmatic selection criterion. This value lies well above the random baseline (0.5 for a balanced binary classification problem) and ensures that only models with clear predictive power are retained, without being overly restrictive. 

A total of 12 classifiers satisfied this condition, and their performance metrics are presented in Table~\ref{tab:classifier_metrics} of Appendix~\ref{app:classifiers}.

\subsection{Voting Classifier}

The final model was constructed using a VotingClassifier~\cite{kuncheva2014combining}, an ensemble learning technique that integrates multiple base classifiers to enhance overall performance by leveraging their complementary strengths. This approach mitigates individual classifier weaknesses and improves generalization. 

\begin{figure}
    \centering
    \begin{subfigure}{0.48\textwidth}
        \centering
        \includegraphics[width=\linewidth]{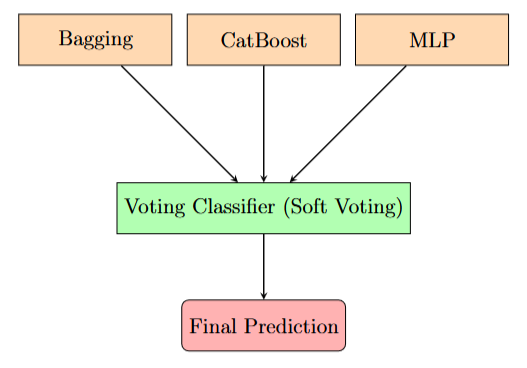}
        \caption{}
    \end{subfigure}
    \begin{subfigure}{0.48\textwidth}
        \centering
        \includegraphics[width=\linewidth]{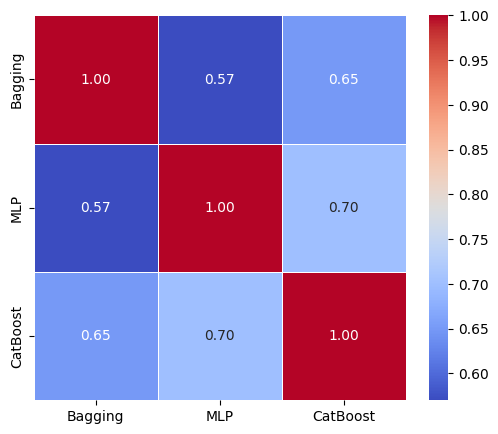}
        \caption{}
    \end{subfigure}
    \caption{Ensemble classifier structure. Final classification is determined by the weighted average of the class probabilities predicted by each base model. Lower inter classifier correlation reduces prediction redundancy, enhancing ensemble effectiveness.}
    \label{fig:voting_structure}
\end{figure}

The ensemble combined three classifiers based on different learning strategies: a BaggingClassifier for variance reduction \cite{breiman1996bagging}, a CatBoostClassifier for efficient gradient boosting \cite{prokhorenkova2018catboost}, and a Multi-Layer Perceptron (MLP) for capturing non-linear patterns \cite{goodfellow2016deep}. This selection was guided by both their individual performance metrics and the diversity of their predictions. As shown in Fig.~\ref{fig:voting_structure} (right), the low inter-correlation among the base classifiers confirms their complementary nature, which reduces prediction redundancy. Combining models with decorrelated errors enhances the ensemble's generalization capability and robustness against overfitting.

Hyperparameter optimization for each base model was performed using RandomizedSearchCV (100 iterations, 5-fold cross-validation), with the optimal parameters detailed in Appendix~\ref{app:hyperparams}. The individual F1-scores achieved on the validation folds were 0.808 for Bagging, 0.813 for CatBoost, and 0.812 for the MLP, which formed the basis for their weights in the ensemble. The ensemble system combines predictions through optimized soft voting. Final class probabilities are computed as:

\begin{equation}
P_{\text{ensemble}}(y=1) = \underbrace{0.3318}_{\text{Bagging weight}} \times P_{\text{Bagging}} + \underbrace{0.3342}_{\text{CatBoost weight}} \times P_{\text{CatBoost}} + \underbrace{0.3340}_{\text{MLP weight}} \times P_{\text{MLP}}
\end{equation}

The numerical coefficients represent the computed weights for each base classifier (Bagging, CatBoost, and MLP, respectively), derived from their individual F1 scores. Each weight is proportional to the classifier's performance, ensuring that models with higher F1 scores contribute more to the ensemble. The weights are normalized to sum to 1.0, reflecting their relative influence. Binary predictions are obtained by applying a threshold of 0.5 to the weighted probability output.

\begin{figure}
    \centering
    \begin{subfigure}{0.48\textwidth}
        \centering
        \includegraphics[width=\linewidth]{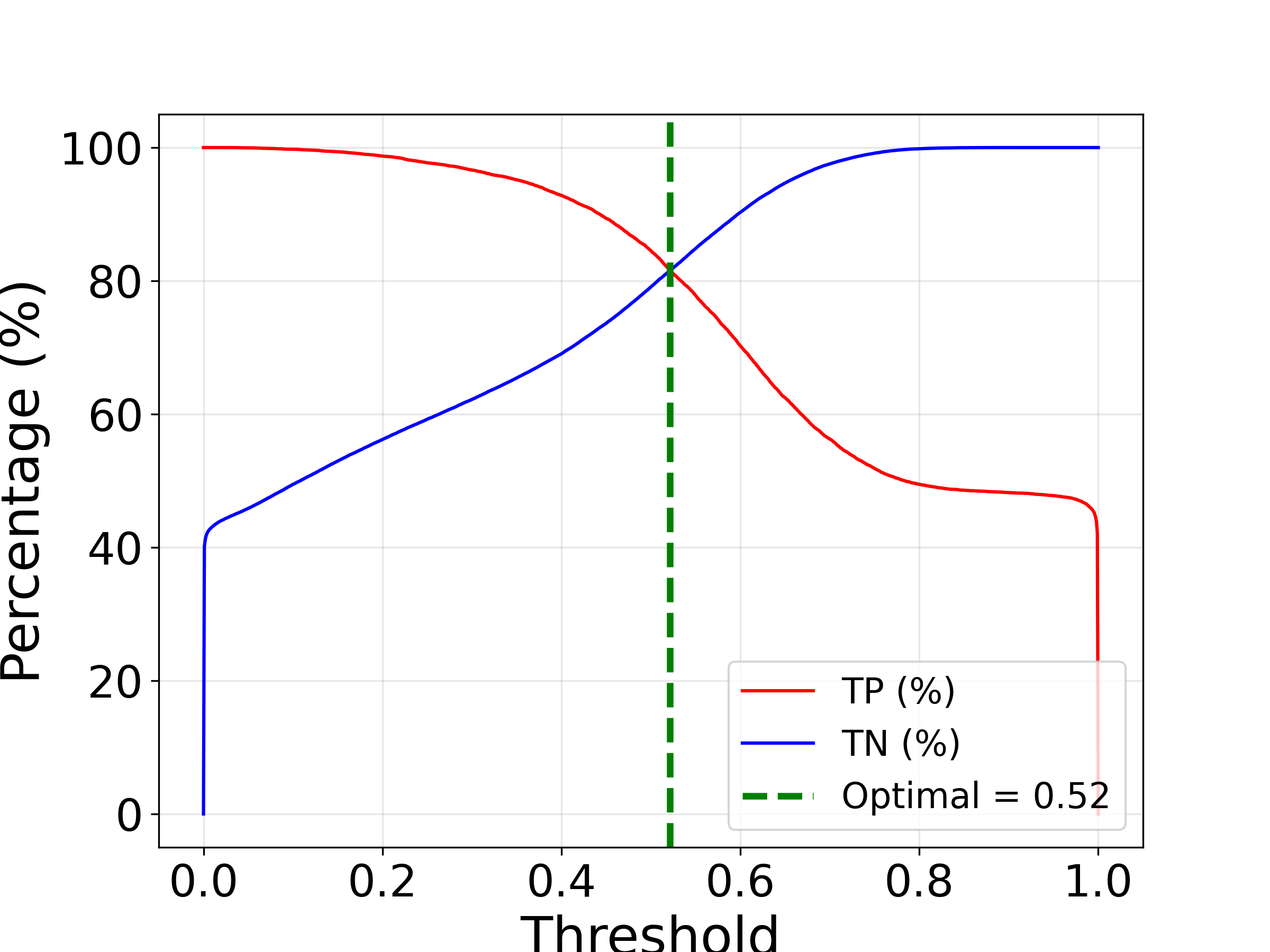}
        \caption{}
    \end{subfigure}
    \begin{subfigure}{0.48\textwidth}
        \centering
        \includegraphics[width=\linewidth]{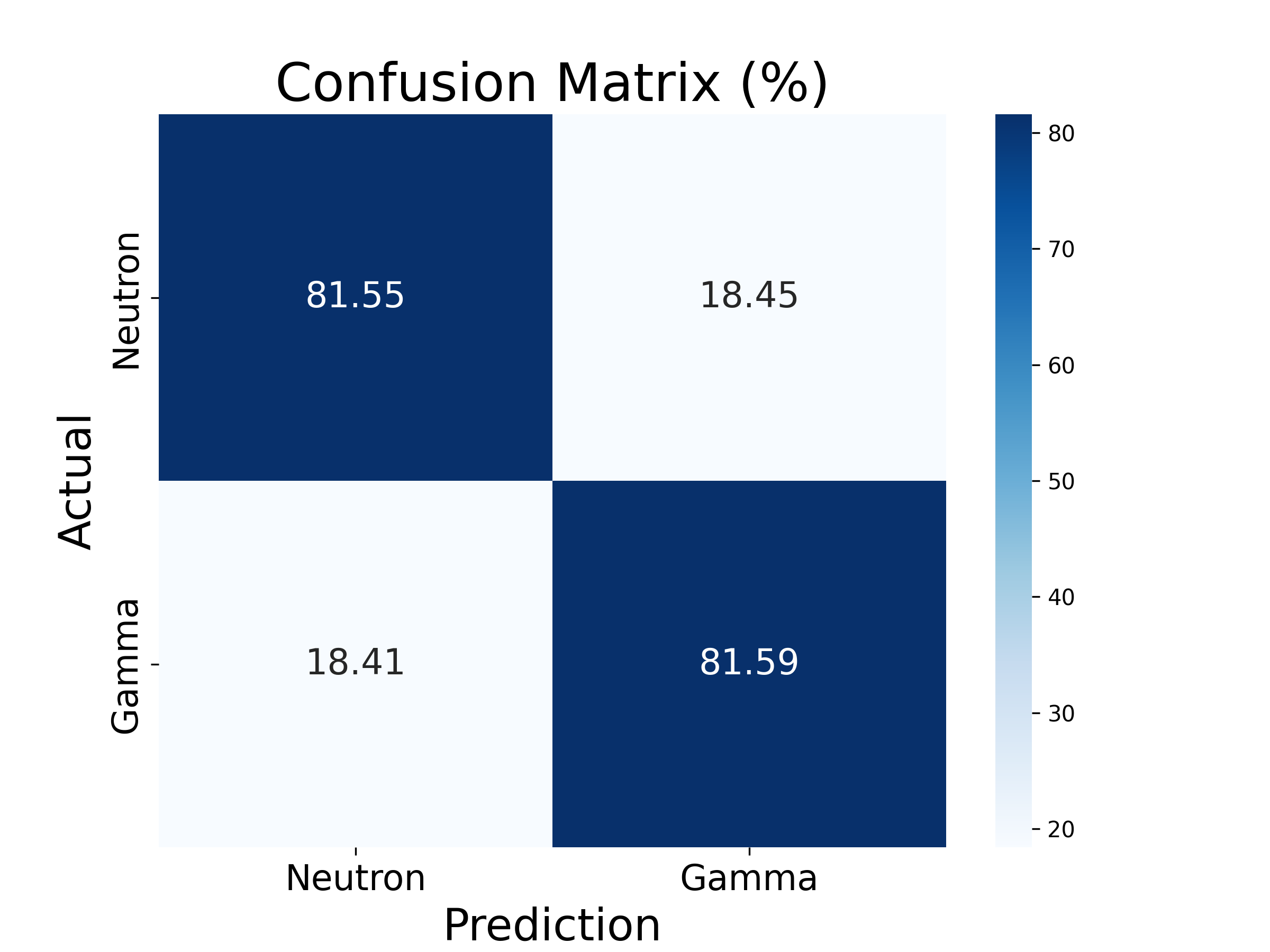}
        \caption{}
    \end{subfigure}
    \caption{Model threshold optimization: The optimal threshold is selected at the intersection point, ensuring a balanced classification of both neutron and gamma signals. The confusion matrix (right) highlights TP, FP, FN and TN after threshold adjustment (0.52). Final test accuracy reaches 0.816.}
    \label{fig:conf_matrix}
\end{figure}

However, to ensure optimal classification performance for both neutron and gamma signals, it is necessary to fine-tune this 0.5 threshold. To determine the most suitable one, the trained ensemble was evaluated on the test set across a range of decision thresholds. The optimal threshold was selected at the point where TP and TN rates were balanced, ensuring that neither class was disproportionately favored. As shown in Fig.~\ref{fig:conf_matrix} left, the optimal decision threshold was determined to be 0.52. This adjustment ensures a balanced and unbiased classification, preventing any preference towards one class over the other.

After establishing the optimal threshold, the confusion matrix was generated to further assess the model's performance. This matrix offers a detailed overview of classification errors and correct predictions. In particular, TP (True Positives) refers to neutrons correctly identified by the model, TN (True Negatives) represents gammas correctly identified, FP (False Positives) indicates gammas misclassified as neutrons, and FN (False Negatives) refers to neutrons misclassified as gammas. As seen in Fig.~\ref{fig:conf_matrix} right, the accuracy of the VotingClassifier model is \(\sim 0.816\).

\begin{figure}
    \centering
    \includegraphics[width=\linewidth]{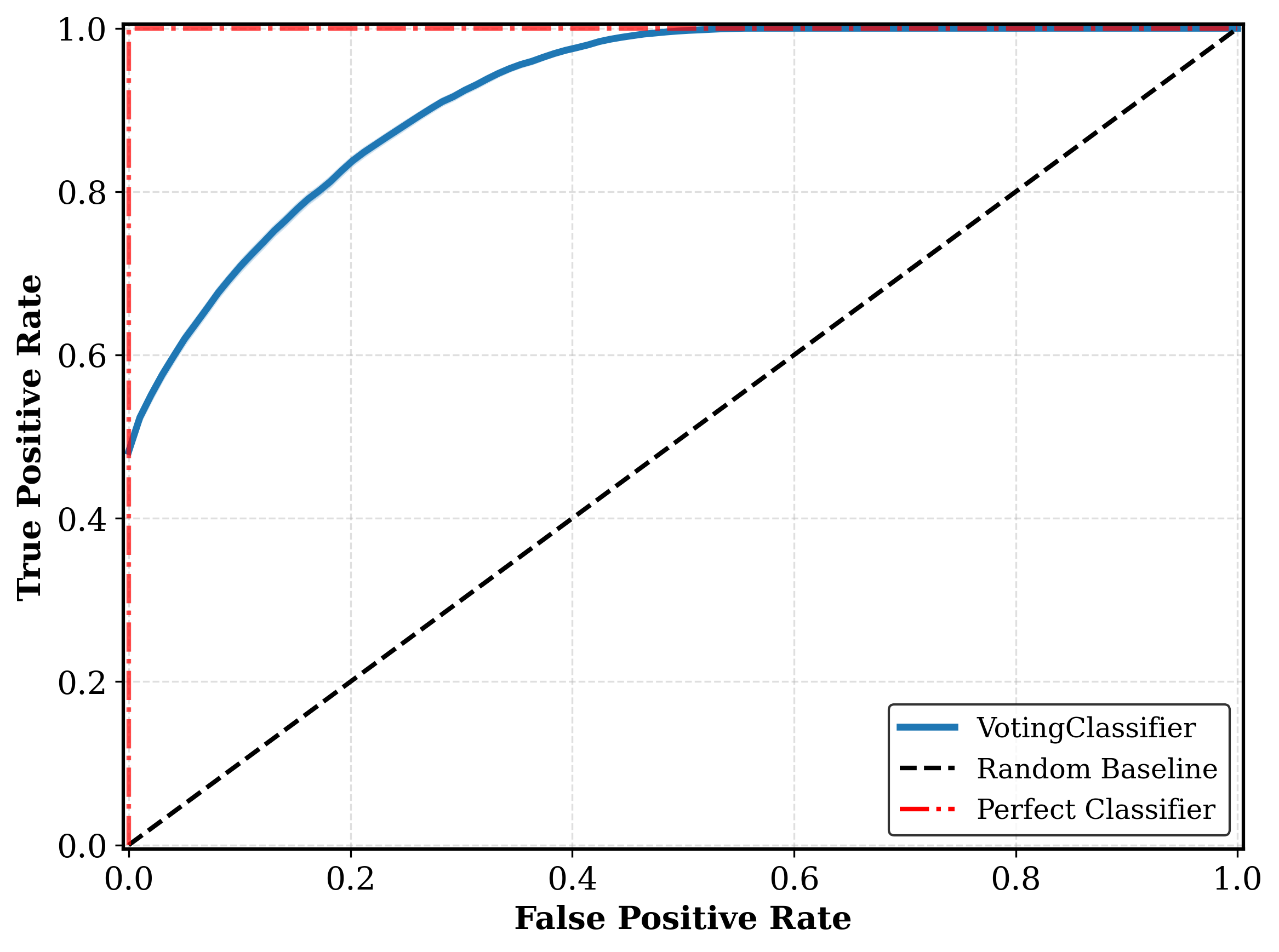}
    \caption{ROC curve showing performance of the Voting Classifier AUC = 0.921 (95\% CI: 0.919-0.922). Dashed lines indicate theoretical performance bounds for Random Baseline (AUC = 0.5) and Perfect Classifier (AUC = 1).}
    \label{fig:roc_curve}
\end{figure}

To further evaluate classification performance across different thresholds, the Receiver Operating Characteristic (ROC) curve was analyzed. The ensemble achieved ROC-AUC=0.921 (95\% CI: 0.919-0.922) through 1000 bootstrap iterations (Fig.~\ref{fig:roc_curve}), signifying strong class separation capability.

Having validated the classifier's performance on the test set, we then applied it to an independent dataset: pulses from the unshielded \(^{241}\)AmBe source, which were not seen during training. This inference step allows us to decompose the mixed radiation field into its neutron and gamma components. Figure~\ref{fig:ambe_predicted} in Section~\ref{sec:2stage} shows the resulting predicted spectra, demonstrating the model's capability to perform a data-driven separation of mixed radiation fields in a real world scenario.

It is important to note that the trained classifier is most effective in the mid energy region (1.5--2.5 MeV), where the natural background count rate is low (Fig.~\ref{fig:twostage}). Although the model performs well in this range (Fig.~\ref{fig:ambe_predicted}), its applicability is constrained by the chosen charge window (4500--8000 ADU), which was selected to maximize signal dominance while minimizing background contamination. This window restricts the method to sources whose signal emerges above background in the high energy, low background region. For low energy gamma sources, even when the signal clearly exceeds background, the elevated background count rate in that spectral region leads to increased signal contamination, degrading classification performance. Extending the approach to lower energies would require either a more sophisticated model capable of handling higher background levels or a preliminary stage that discriminates and subtracts natural background components. Future work will explore these strategies to broaden the operational energy range of the classifier.

\section{Conclusions}
\label{sec5}

This work has presented and validated an integrated methodology for neutron-gamma discrimination in a pure water Cherenkov detector, combining statistically grounded spectral analysis with pulse level machine learning classification. The proposed framework operates as a coherent two-stage workflow that integrates the interpretability of physical detection principles alongside the discriminative power of data driven algorithms, thereby extending the analytical capability of water Cherenkov detectors beyond their traditional applications in high energy physics and cosmic ray studies.

The first stage introduces a statistically principled method for threshold determination based on a $3\sigma$ significance criterion, which identifies the charge cutoff at which source spectra become statistically indistinguishable from background. This approach, detailed in Section~\ref{sec3}, yields objective and reproducible energy thresholds that enable rapid screening of low energy gamma emitters while substantially reducing the volume of data requiring computationally intensive pulse level analysis. The resulting cutoff values, converted to energy equivalent thresholds via the linear calibration relation established in Eq.~\ref{eq:fit_inverse}, provide a physically interpretable first order classification that leverages the detector's intrinsic response characteristics.

For sources whose spectra extend beyond the neutron threshold region (the green band in Fig.~\ref{fig:twostage}), a second stage employs supervised machine learning to achieve pulse level discrimination between neutron and gamma signals. As described in Section~\ref{sec:ml}, multiple algorithms were evaluated, with a weighted soft-voting ensemble combining Bagging, CatBoost, and a Multi-Layer Perceptron yielding the strongest performance. The ensemble classifier achieved an accuracy of approximately 0.816 and a ROC-AUC of 0.921, the confusion matrix indicates balanced classification performance across both radiation types, Fig.~\ref{fig:conf_matrix}, while the ROC curve demonstrates strong class separability, Fig.~\ref{fig:roc_curve}. Following validation on the test set, the classifier was successfully applied to an independent dataset comprising unshielded \(^{241}\)AmBe pulses, effectively decomposing the mixed radiation field into its constituent neutron and gamma components, as shown in Fig.~\ref{fig:ambe_predicted}.

The integration of both stages into a unified decision workflow, illustrated in Fig.~\ref{fig:workflow}, constitutes the principal methodological contribution of this work. This hybrid architecture integrates the complementary strengths of statistical and data driven approaches: the former provides physical interpretability and computational efficiency for initial screening, while the latter enhances discrimination capability through learned waveform features. 

An important consideration for the operational deployment of this framework is the long term stability of the detector response. In our approach, the first stage defines the neutron threshold via energy calibration; therefore, any significant drift in PMT gain or aging effects would alter the ADU to MeV conversion and shift the energy window used for pulse level classification. During the data acquisition period of this study, the detector response remained stable and no gain variations were observed. However, in practice, a change in detector response would require recalibration, which would redefine the region of interest. If the recalibration leads to a substantial redistribution of pulse characteristics within the ML domain, retraining of the classifier would be performed as part of the standard calibration procedure. 

A dedicated long term study of classifier robustness under controlled gain variations is planned for future work, alongside dedicated Monte Carlo simulations to investigate the precise event topology of multiple Compton scatters, optical photon propagation, and their contribution to the observed pulse shape variations. Additionally, further research will focus on extending this framework toward lower energy spectra, aiming to develop statistical and machine learning models capable of discriminating source types in both shielded and unshielded configurations while effectively removing background contributions prior to analysis. Such advancements would further generalize the methodology and broaden its applicability to operational radiation detection deployments.

In summary, this study validates a robust, scalable, and operationally feasible framework for radiation source discrimination in water Cherenkov detectors. By combining classical significance-based statistical methods with modern ensemble learning techniques, the analytical capability of these detectors is substantially enhanced, positioning them as versatile instruments for both fundamental research and applied radiation detection.

\section*{Acknowledgements}

The authors acknowledge co-funding from the Programa Iberoamericano de Ciencia y Tecnología para el Desarrollo (CYTED) through the LAGO-INDICA network (Project 524RT0159-LAGO-INDICA: Infraestructura digital de ciencia abierta). 

This work was partially supported by the CLAF-HECAP Programme through a research mobility scholarship awarded to Christian Sarmiento-Cano for a visit to the Neutron Physics Department at Centro Atómico Bariloche.

AI technology was used to proofread and polish this manuscript. After using this tool, the authors reviewed and edited the content as needed and take full responsibility for the content of the published article.

\appendix

\section{Uncertainty Estimation of the Cutoff Point}
\label{app:cutoff_uncertainty}

The total standard uncertainty in the cutoff position, \(\sigma_{\mathrm{cutoff}}\), accounts for statistical fluctuations, methodological sensitivity to the selected significance criterion, and discretization effects due to finite bin width.

\subsection{Statistical Uncertainty}

The statistical component (\(\sigma_{\mathrm{stat}}\)) was estimated via bootstrap resampling with \(N=1000\) realizations. In each iteration, the source and background histograms were independently re-sampled assuming Poisson statistics with means equal to the observed counts. The cutoff was recalculated for each realization, and the standard deviation of the resulting distribution defines \(\sigma_{\mathrm{stat}}\) \cite{davison1997bootstrap}.

\subsection{Threshold Sensitivity}

To quantify the dependence of the cutoff on the selected significance criterion, the procedure was repeated using \(1\sigma\) and \(5\sigma\) compatibility thresholds. The associated uncertainty is defined as

\begin{equation}
\sigma_{\mathrm{thr}} =
\frac{1}{2}\left| \mathrm{CP}_{5\sigma} - \mathrm{CP}_{1\sigma} \right|.
\end{equation}

This term reflects the methodological sensitivity of the estimator to the chosen compatibility threshold.

\subsection{Discretization Uncertainty}

The discretization uncertainty is derived from the standard deviation of a uniform distribution over a finite bin width \(w_{\mathrm{bin}}\). Under this assumption, the variance is given by \(w_{\mathrm{bin}}^2/12\), leading to \(\sigma_{\mathrm{bin}} = w_{\mathrm{bin}}/\sqrt{12}\) \cite{bevington2003data}. This corresponds to the standard treatment of quantization uncertainty in binned measurements.

\subsection{Total Uncertainty}

The total cutoff uncertainty is obtained by quadratic combination:

\begin{equation}
\sigma_{\mathrm{cutoff}} =
\sqrt{
\sigma_{\mathrm{stat}}^2 +
\sigma_{\mathrm{thr}}^2 +
\sigma_{\mathrm{bin}}^2
}.
\label{eq:sigma_cutoff_total}
\end{equation}

The final calibration point is reported as

\begin{equation}
\mathrm{CP} = \mathrm{cutoff\ bin} \pm \sigma_{\mathrm{cutoff}}.
\end{equation}

In order to quantify the relative contribution of each uncertainty source, the different terms entering Eq.~\ref{eq:sigma_cutoff_total} were evaluated for all calibration datasets. The results show that the dominant contribution arises from the threshold sensitivity term, \(\sigma_{\mathrm{thr}}\), which typically accounts for \(\sim 70\)--\(80\%\) of the total uncertainty. The statistical component, \(\sigma_{\mathrm{stat}}\), contributes at the level of \(\sim 20\%\), while the discretization term, \(\sigma_{\mathrm{bin}}\), remains subdominant (\(\lesssim 10\%\)).

This hierarchy reflects the fact that the cutoff determination is primarily limited by the methodological choice of the significance criterion, rather than by counting statistics or binning effects.

\section{Supplementary Classification Results}

\subsection{Base classifiers metrics}
\label{app:classifiers}

\begin{table}
\centering
\adjustbox{max width=\linewidth}{%
\begin{tabular}{l c c c c c}
\toprule
\textbf{Classifier} & \textbf{Accuracy} & \textbf{Precision} & \textbf{Recall} & \textbf{F1-Score} & \textbf{Log Loss} \\
\midrule
RandomForest         & 0.795 & 0.876 & 0.795 & 0.818 & 0.336 \\
HistGradientBoosting & 0.777 & 0.878 & 0.777 & 0.805 & 0.327 \\
ExtraTrees           & 0.781 & 0.873 & 0.781 & 0.807 & 0.377 \\
Bagging              & 0.815 & 0.868 & 0.815 & 0.832 & 0.434 \\
AdaBoost             & 0.772 & 0.872 & 0.772 & 0.800 & 0.550 \\
XGBoost              & 0.784 & 0.877 & 0.784 & 0.809 & 0.327 \\
LightGBM             & 0.776 & 0.878 & 0.776 & 0.803 & 0.328 \\
CatBoost             & 0.794 & 0.880 & 0.794 & 0.818 & 0.323 \\
SVC\_RBF             & 0.768 & 0.869 & 0.768 & 0.796 & {--} \\
NuSVC                & 0.766 & 0.867 & 0.766 & 0.794 & {--} \\
MLP                  & 0.798 & 0.879 & 0.798 & 0.821 & 0.336 \\
MLP\_Deep            & 0.792 & 0.873 & 0.792 & 0.816 & 0.380 \\
\bottomrule
\end{tabular}%
}
\caption{Classification performance metrics across different classifiers (Accuracy $>$ 0.75). Log Loss: lower is better, -- indicates non probabilistic classifiers.}
\label{tab:classifier_metrics}
\end{table}

\subsection{Optimized Hyperparameters}
\label{app:hyperparams}

The following hyperparameters were obtained via RandomizedSearchCV (100 iterations, 5-fold cross-validation):

\begin{verbatim}
BaggingClassifier: {estimator: DecisionTreeClassifier, bootstrap: True,
 n_estimators: 200, max_samples: 1.0, max_features: 1.0,
 estimator__min_samples_split: 2, estimator__min_samples_leaf: 1,
 estimator__max_depth: None} (F1-Score: 0.808)

CatBoostClassifier: {random_strength: 1.0, learning_rate: 0.05,
 l2_leaf_reg: 5, iterations: 1500, depth: 6, border_count: 128,
 bagging_temperature: 1} (F1-Score: 0.813)

MLP: {solver: 'sgd', max_iter: 500, learning_rate_init: 0.01,
 learning_rate: 'adaptive', hidden_layer_sizes: (100, 50, 25),
 early_stopping: True, alpha: 0.1, activation: 'relu'} (F1-Score: 0.812)
\end{verbatim}

\bibliographystyle{plain}
\bibliography{bibliography}

\end{document}